\documentclass{emulateapj}

\shorttitle{EELR and the AGN Feedback}
\shortauthors{Matsuoka}

\begin{document}


\title{Co-evolution of Galaxies and Central Black Holes: \\Observational Evidence on the Trigger of AGN Feedback}



\author{Y. Matsuoka}
\affil{Graduate School of Science, Nagoya University, Furo-cho, Chikusa-ku, Nagoya 464-8602, Japan.}
\email{matsuoka@a.phys.nagoya-u.ac.jp}


\begin{abstract}
A comprehensive analysis of extended emission line region (EELR) around quasars is presented.
New Subaru/Suprime-Cam observation is combined with literature search, resulting in a compilation of 81 EELR measurements 
for type-1 and type-2 quasars with associated active galactic nucleus (AGN) and host galaxy properties.
It is found that EELR phenomenon shows clear correlation with Eddington ratio, which links EELR to the constituents of the principal component 1 (PC 1), 
or eigenvector 1, of the AGN emission correlations.
We also find that EELR is preferentially associated with gas-rich, massive blue galaxies.
It supports the idea that the primary determinant of EELR creation is the gas availability and that the gas may be brought in by galaxy
merger triggering the current star formation as well as AGN activity, and also gives an explanation for the fact that
most luminous EELR is found around radio-loud sources with low Eddington ratio.
By combining all the observations, it is suggested that EELR quasars occupy the massive blue corner of the green valley, the AGN realm,
on the galaxy color - stellar mass diagram.
Once a galaxy is pushed to this corner, activated AGN would create EELR by the energy injection into the interstellar gas and eventually blow 
it away, leading to star-formation quenching.
The results presented here provide a piece of evidence for the presence of such AGN feedback process, which may be playing a leading role in 
the co-evolution of galaxies and central super-massive black holes.
\end{abstract}





\section{Introduction \label{sec:intro}}

Some active galactic nuclei (AGNs) are surrounded by massive ionized nebulae extending to several tens of kiloparsec from the central engines. 
The first systematic study of such extended emission-line region (EELR) around quasars in combination with nuclear spectral properties
was presented by \citet{boroson84} and \citet{boroson85}.
They found that nearly half of their 24 objects are EELR quasars \citep[quasars with detectable EELR, following][]{fu07} and that the EELR
quasars have larger [\ion{O}{3}] $\lambda$5007 equivalent width (EW), broader and bumpier Balmer lines, and weaker \ion{Fe}{2} emission
compared to non-EELR quasars.
In addition, the EELR quasars are predominantly associated with steep-spectrum radio emission with extended, lobe-dominant morphology, 
in contrast to the non-EELR quasars associated with flat-spectrum radio emission with compact, core-dominant morphology or without strong
radio emission.

The correlation between EELR phenomenon and radio emission was further confirmed by \citet{stockton87}, who obtained narrow-band [\ion{O}{3}] line images of 47 quasars.
Above the limiting luminosity of extended [\ion{O}{3}] emission $L_{\rm E[O III]} = 5 \times 10^{41}$ erg s$^{-1}$, 
10 out of 26 (10/26) of their steep-spectrum radio quasars (SSRQs) have EELR, while majority of the flat-spectrum radio 
quasars (FSRQs) and the radio-quiet quasars (RQQs; 7/7 of the FSRQs and 13/14 of the RQQs) are non-EELR sources \citep{fu09}.
They also found a correlation between nuclear [\ion{O}{3}] luminosity and the presence of EELR.
\citet{fu07} analyzed the rest-frame ultraviolet spectra of 12 SSRQs in the \citet{stockton87} sample and found that \ion{N}{5} $\lambda$1240 emission is suppressed
in EELR quasars, which they claimed points to lower metallicity of the broad-line region (BLR) compared to non-EELR quasars.
They did not find any systematic difference in other AGN properties such as continuum luminosity and mass of super-massive black holes (SMBHs)
between the EELR and non-EELR SSRQs.

On the other hand, \citet{husemann08} observed 20 quasars dominated by RQQs and found EELR around 8 of them.
The correlation between nuclear \ion{Fe}{2} EW, H$\beta$ line width, and EELR was confirmed in their sample too.
In addition, they found a dependence of EELR on SMBH mass ($M_{\rm BH}$) and reported that $M_{\rm BH} = 10^{8.5} M_{\odot}$ separates EELR quasars 
with larger $M_{\rm BH}$ from non-EELR quasars with smaller $M_{\rm BH}$.
Searches for EELR around RQQs were also presented by \citet{humphrey10} and \citet{villar-martin10,villar-martin11}
focusing on obscured, type-2 quasars. 
They found EELR around 10 out of 20 objects and claimed that luminous EELR is not necessarily associated with powerful radio sources.

Observational constraints on the emergence of EELR can possibly be a key to understand the co-evolution of galaxies and SMBHs.
It is now widely accepted that the tight correlation between SMBH mass and bulge mass of their host galaxies \citep[e.g.,][]{magorrian98, haring04} 
indicates the intimately-connected evolution of the two systems.
The most compelling process of interaction between galaxies and SMBHs is the AGN feedback, in which the energy emitted by AGN regulates the mass accretion 
to SMBHs and drives galactic wind that can expel cold gas from and quench star formation in host galaxies.
The AGN feedback could also give a solution to the two major problems in the current galaxy formation models based on the $\Lambda$ cold dark matter (CDM) theory: 
first is the "overcooling" problem in which
much more massive galaxies are formed in the numerical simulations than observed due to too efficient gas cooling, and second is the
"inverted color - mass - morphology relation" problem in which massive galaxies are predicted to be blue and disk-dominant rather than red and
spheroid-dominant as observed \citep[e.g.,][]{somerville08}.
However, observational evidence is still lacking on whether the AGN feedback is actually present or not and how it works if present.
While measurements of galaxy evolution imprinted on the stellar-mass (luminosity) function \citep[e.g.,][]{bell04,faber07,brown07,brown08} especially at
its massive end \citep{conselice07,matsuoka10,matsuoka11a} and its integration \citep[the extragalactic background light; e.g, ][]{matsuoka11b},
or analyses of the nature of the AGN phenomenon \citep[e.g.,][]{laor00,boroson02,matsuoka07,matsuoka08,vestergaard08} can put indirect constraints on the co-evolution, 
EELR possibly provides more direct clues as to the interrelation between galaxies and SMBHs.
The previous measurements indicate that EELR gas is largely ionized by AGN rather than stellar population \citep{fu09,husemann10,keel11}, 
hence we may be witnessing the AGN feedback in action in this galaxy-wide ionization process.
Indeed, \citet{greene11} recently showed that AGN photoionizes and kinematically disturbs the interstellar medium throughout the entire host galaxies
of luminous type-2 quasars.

In this paper, new observation of EELR around five quasars at $z \sim 0.3$ is first presented.
Then it is combined with the previous observations of EELR and associated AGN and host galaxy properties obtained mainly from 
the Sloan Digital Sky Survey \citep[SDSS;][]{york00} archive.
By using this compiled sample as well as the control samples of SDSS quasars and galaxies, we aim to perform a comprehensive analysis of EELR in the context of 
galaxy evolution.
A cosmology with $H_0$ = 70 km s$^{-1}$ Mpc$^{-1}$, $\Omega_{\rm M} = 0.3$, and $\Omega_{\rm \Lambda} = 0.7$ is assumed throughout this work.
Luminosity and mass are given in units of erg s$^{-1}$ and the solar mass ($M_{\odot}$) unless otherwise noted.
All the presented magnitudes are on the AB system.

\section{Data}


\subsection{New Observation}

The observed quasars were selected from the SDSS Data Release 7 (DR7) quasar catalog \citep{schneider10,shen11}.
The simple selection criteria were set based on redshift ($z \simeq 0.3$ in order that [\ion{O}{3}] $\lambda$5007 falls in the narrow-band filter transmission
described below) and nuclear [\ion{O}{3}] luminosity ($L_{\rm N[O III]} > 10^{42}$ erg s$^{-1}$).
Among the objects meeting the criteria, those with various SMBH mass and radio loudness and with the highest [\ion{O}{3}] luminosity were picked up.
The properties of the observed quasars are summarized in Table \ref{tab:objprop}.
They effectively complement the existing measurements on the AGN principal-component plane explored below.

The targets were observed with the prime-focus imaging facility Suprime-Cam \citep{miyazaki02} mounted on Subaru telescope\footnote{
This work is based in part on data collected at Subaru Telescope, which is operated by the National Astronomical Observatory of Japan (NAOJ).}.
The instrument covers a wide field-of-view of 34 $\times$ 27 arcmin$^2$ with ten 2k $\times$ 4k CCDs, with a pixel scale of 0.20 arcsec.
The narrow-band NA$_{656}$ (effective wavelength $\lambda_{\rm eff}$ = 6570 \AA, effective bandpass $\Delta\lambda_{\rm eff}$ = 140 \AA)
and the broad-band $R_{\rm c}$ ($\lambda_{\rm eff}$ = 6530 \AA, $\Delta\lambda_{\rm eff}$ = 1110 \AA) filters were used to trace the redshifted 
[\ion{O}{3}] line and underlying continuum, respectively.
The observation was carried out on 2011 May 3--4 under the open-use program S11A-028.
The sky condition was sometimes non-photometric, but the measured flux of celestial objects was mostly stable within 10 \% variation.
The seeing was 0.8--1.1 arcsec.
For each target, nine 300-sec or three 900-sec exposures in NA$_{656}$ and nine 100-sec or three 300-sec exposures in $R_{\rm c}$ were obtained,
depending on the source brightness.
In addition, a spectrophotometric standard star SA 104-335 was observed with the same instrument configuration as the target observation.

\begin{table*}
\begin{center}
\caption{Targets of Subaru/Suprime-Cam observation.\label{tab:objprop}}
\begin{tabular}{ccccccccc}
\tableline\tableline
                 & $r$-band  &      & log $L_{\rm N[O\ III]}$ & log $M_{\rm BH}$ &    & log $L_{\rm E[O\ III]}$ \\
Quasar           & mag.\tablenotemark{a} & Redshift & (erg s$^{-1}$)  & ($M_{\odot}$) & R\tablenotemark{b} & (erg s$^{-1}$)\\
\tableline
SDSS 091401.75$+$050750.6 & 17.32 & 0.301   & 42.85 $\pm$ 0.01                       & 9.36 $\pm$ 0.10 & L & 42.17 $\pm$ 0.08 \\ 
SDSS 095456.89$+$092955.7 & 17.92 & 0.298   & 42.93 $\pm$ 0.01                       & 8.56 $\pm$ 0.08 & L & $<$ 41.47        \\ 
SDSS 113949.47$+$402048.5 & 17.52 & 0.314   & 42.97 $\pm$ 0.01                       & 8.21 $\pm$ 0.08 & Q & $<$ 41.48        \\ 
SDSS 150740.92$+$445331.5 & 18.16 & 0.314   & 42.69 $\pm$ 0.01                       & 7.34 $\pm$ 0.14 & L & $<$ 41.53        \\ 
SDSS 150752.66$+$133844.5 & 17.79 & 0.322   & 42.53 $\pm$ 0.01                       & 9.67 $\pm$ 0.02 & Q & 41.83 $\pm$ 0.08 \\ 
\tableline
\end{tabular}
\end{center}
Note --- $^{\rm a}$SDSS PSF magnitude. The measurement error is $<$ 0.02 mag.
$^{\rm b}$Radio loudness. Q: radio quiet, L: radio loud.
\end{table*}

Data reduction was performed with the Suprime-cam Deep Field Reduction \citep[SDFRED;][]{ouchi04} software in a standard manner, including
bias subtraction, flat fielding, distortion correction, sky subtraction, masking bad pixels, and stacking.
Photometric calibration in NA$_{656}$ was achieved by referring to the observed flux of SA 104-335, whose intrinsic magnitude
was derived by convolving its spectrophotometry data \citep{stone96} with the filter transmission function.
Then the $R_{\rm c}$ images were calibrated relative to NA$_{656}$ by requiring the mean ($R_{\rm c}$ $-$ NA$_{\rm 656}$) color to be zero 
for galaxies detected with high signal-to-noise ratios (point sources were not used because stellar H$\alpha$ absorption could affect this calibration).
The scatter of $\sim$0.05 mag was found around the mean between $R_{\rm c }$ and NA$_{\rm 656}$ magnitudes in the calibration.

\begin{figure}
\epsscale{1.1}
\plotone{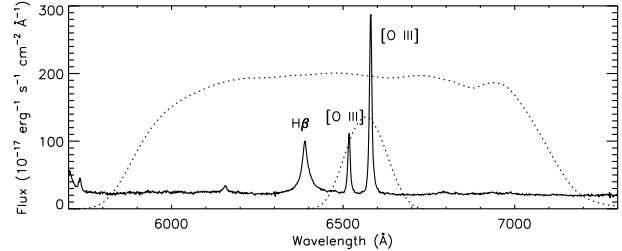}
\caption{Nuclear spectrum of SDSS 1139$+$4020 (solid line) with the arbitrary-scaled transmission functions of broad-band $R_{\rm c}$ 
  and narrow-band NA$_{656}$ filters (dotted lines).\label{sdssspec}}
\end{figure}

The reduction process then proceeded to the continuum subtraction from the NA$_{656}$ images.
For this purpose, the nuclear spectra of the quasars were retrieved from the SDSS DR7 archive\footnote{
Funding for the SDSS and SDSS-II has been provided by the Alfred P. Sloan Foundation, the Participating Institutions, the National Science Foundation, 
the U.S. Department of Energy, the National Aeronautics and Space Administration, the Japanese Monbukagakusho, the Max Planck Society, and the Higher 
Education Funding Council for England. The SDSS Web Site is http://www.sdss.org/.
The SDSS is managed by the Astrophysical Research Consortium for the Participating Institutions. The Participating Institutions are the American Museum 
of Natural History, Astrophysical Institute Potsdam, University of Basel, University of Cambridge, Case Western Reserve University, University of Chicago, 
Drexel University, Fermilab, the Institute for Advanced Study, the Japan Participation Group, Johns Hopkins University, the Joint Institute for Nuclear 
Astrophysics, the Kavli Institute for Particle Astrophysics and Cosmology, the Korean Scientist Group, the Chinese Academy of Sciences (LAMOST), Los Alamos 
National Laboratory, the Max-Planck-Institute for Astronomy (MPIA), the Max-Planck-Institute for Astrophysics (MPA), New Mexico State University, Ohio State 
University, University of Pittsburgh, University of Portsmouth, Princeton University, the United States Naval Observatory, and the University of Washington.}
(an example is shown in Figure \ref{sdssspec}).
They were measured within SDSS spectroscopic fibers of 3-arcsec aperture.
For each spectrum, the continuum levels beneath [\ion{O}{3}] $\lambda$4959 and $\lambda$5007 were estimated by interpolation between continuum windows at
both (shorter- and longer-wavelength) sides of the lines, and then the spectrum was decomposed into [\ion{O}{3}] and continuum.
Convolving them with the filter transmissions gives us the relative fluxes of [\ion{O}{3}] and continuum in NA$_{656}$
($f_{\rm NA_{656}}^{\rm [O III]}$ and $f_{\rm NA_{656}}^{\rm cont}$) and in $R_{\rm c}$ ($f_{R_{\rm c}}^{\rm [O III]}$ and $f_{R_{\rm c}}^{\rm cont}$) 
for each object.
Using these measures, accurate continuum subtraction was achieved as follows.
The radial profiles of all the quasar nuclei on the obtained images are consistent with the point spread functions (PSFs) derived from nearby stars,
hence they are dominated by the radiation from central unresolved sources.
The NA$_{656}$ and $R_{\rm c}$ images were decomposed into nuclear (${\rm NA}_{656}^{\rm nuc}$, ${R}_{\rm c}^{\rm nuc}$) and extended 
(${\rm NA}_{656}^{\rm ext}$, ${R}_{\rm c}^{\rm ext}$) components by the best-fit Moffat profiles, fitted within a 10-pixel (2-arcsec) aperture
on top of the quasar peak positions.
Then the nuclear [\ion{O}{3}] distributions are given by
\begin{equation}
  {\rm NA}_{656}^{\rm nuc} \times \frac{f_{\rm NA_{656}}^{\rm [O III]}}{f_{\rm NA_{656}}^{\rm [O III]} + f_{\rm NA_{656}}^{\rm cont}} ,
  \label{eq:contsub_nuc}
\end{equation}
while the extended [\ion{O}{3}] distributions are
\begin{equation}
  ({\rm NA}_{656}^{\rm ext} -  {R}_{\rm c}^{\rm ext}) \times \frac{f_{\rm NA_{656}}^{\rm [O III]}}{f_{\rm NA_{656}}^{\rm [O III]} - f_{R_{\rm c}}^{\rm [O III]}} .
  \label{eq:contsub_ext}
\end{equation}
The second term in the expression (\ref{eq:contsub_ext}), which is equivalent in principle to the ratio
$\Delta\lambda_{\rm eff, R_{\rm c}}$ / ($\Delta\lambda_{\rm eff, R_{\rm c}} - \Delta\lambda_{\rm eff, NA_{656}}$),
corrects for the continuum oversubtraction of the first term due to the [\ion{O}{3}] themselves contained in ${R}_{\rm c}^{\rm ext}$.
Note that the expression (\ref{eq:contsub_nuc}), although it is most accurate, is not applicable to the extended components since the relative fluxes of
[\ion{O}{3}] and continuum are different from $f_{\rm NA_{656}}^{\rm [O III]}$ and $f_{\rm NA_{656}}^{\rm cont}$ measured in the SDSS spectroscopic fibers.
It is worth re-emphasizing that the spectral information is quite useful in the above process as an accurate cross-calibrator of the images taken
in the different filters.

\subsection{Data Compilation \label{sec:datacomp}}

Thanks to the massive spectroscopic observations conducted by the SDSS, nuclear properties are now known for many quasars 
with EELR measurements in the literature.
We compile the EELR measurements of type-1 quasars from \citet{boroson85}, \citet{stockton87}, and \citet{husemann08,husemann10}.
The continuum and line luminosities given in \citet{stockton87} are converted to the values in the cosmology adopted here.
The measurements of \citet{husemann08} are corrected in accordance with their re-analysis (Husemann et al., in preparation).
Then we cross-match the quasars with the SDSS DR7 quasar catalog \citep{shen11} and obtain the nuclear properties such as continuum and 
line luminosity, SMBH mass, and radio characteristics based on the Faint Images of the Radio Sky at Twenty cm survey \citep[FIRST;][]{becker95}.
For some objects without counterparts in the SDSS catalog, the nuclear properties are taken from the above papers as well as \citet{fu07}.
Combined with the new observation presented in this paper, we end up with 61 type-1 quasars with EELR measurements (27 detections and 34 non-detections) 
and with a variety of nuclear properties.
The compiled sample is listed in Table \ref{tab:nuclear_props}.
We assign radio flag 0 to RQQs, flag 1 to flat-spectrum (radio spectral index $\alpha_\nu < 0.5$) or core-dominant 
\citep[based on the FIRST morphology;][]{jiang07} radio quasars, flag 2 to steep-spectrum ($\alpha_\nu > 0.5$) or lobe-dominant radio quasars.
The SDSS quasars are separated into radio-quiet and radio-loud sources at the radio-loudness parameter $R$ = 5 \citep{shen11}.
For simplicity the objects with radio flag 0, 1, 2 are hereafter referred to as RQQs, FSRQs, and SSRQs, respectively, since there is a well-known, 
strong correlation between the radio morphology and spectral index.

\begin{deluxetable*}{cccccccc}
\tablewidth{0pt}
\tablecaption{Nuclear properties of type-1 quasars with EELR measurements. \label{tab:nuclear_props}}
\tablehead{\colhead{} & \colhead{} & \colhead{log $\lambda L_{\rm \lambda 5100}$}  & \colhead{log $M_{\rm BH}$} & \colhead{} 
  & \colhead{log $L_{\rm N[O III]}$} & \colhead{log $L_{\rm E[O III]}$} &\colhead{}\\
\colhead{Quasar} & \colhead{Redshift} & \colhead{(erg s$^{-1}$)}  & \colhead{($M_{\odot}$)}    & \colhead{R\tablenotemark{a}} & \colhead{(erg s$^{-1}$)} 
& \colhead{(erg s$^{-1}$)} & \colhead{Ref.}}
\startdata
     4C 15.01   & 0.450    &    45.61                        &    9.24             &   2   &    43.58            & $<$ 42.56        & 2  \\
  PG 0026$+$129   & 0.142    &    44.96 $\pm$ 0.11           &    8.98             &   0   &    42.53 $\pm$ 0.09 & 41.80 $\pm$ 0.14 & 3   \\
  PG 0050$+$124   & 0.061    &    44.31 $\pm$ 0.11           &    8.00             &   0   &    42.26 $\pm$ 0.09 & $<$ 43.0         & 3   \\
  PG 0052$+$251   & 0.155    &    44.98 $\pm$ 0.10           &    8.55             &   0   &    42.70 $\pm$ 0.09 & 41.75 $\pm$ 0.13 & 1, 2, 3 \\ 
SDSS 0057$+$1446  & 0.172    &    44.93 $\pm$ 0.01           &    9.38 $\pm$ 0.02  &   0   &    42.23 $\pm$ 0.03 & $<$ 43.0         & 3   \\ 
 HE 0132$-$0441   & 0.154    &    44.69 $\pm$ 0.10           &    7.73             &   0   &    42.15 $\pm$ 0.16 & $<$ 43.0         & 3   \\
  NAB 0137$-$01   & 0.334    &    45.17 $\pm$ 0.01           &    9.01 $\pm$ 0.02  &   0   &    42.62 $\pm$ 0.02 & $<$ 41.53        & 2   \\
SDSS 0155$-$0857  & 0.165    &    44.43 $\pm$ 0.02           &    8.82 $\pm$ 0.04  &   0   &    42.15 $\pm$ 0.02 & 41.43 $\pm$ 0.19 & 3   \\
 HE 0157$-$0406   & 0.218    &    44.72 $\pm$ 0.09           &    8.52             &   0   &    42.09 $\pm$ 0.10 & $<$ 43.0         & 3   \\
  Mrk 1014      & 0.163      &    44.78 $\pm$ 0.01           &    8.06 $\pm$ 0.05  &   1   &    42.74 $\pm$ 0.02 & 42.22 $\pm$ 0.14 & 2, 3   \\
       OI 287   & 0.444      &    45.08 $\pm$ 0.01           &    9.09 $\pm$ 0.21  &   2   &    43.51 $\pm$ 0.01 & $<$ 42.47        & 2   \\
SDSS 0836$+$4426  & 0.254    &    45.29 $\pm$ 0.01           &    8.73 $\pm$ 0.06  &   0   &    43.28 $\pm$ 0.01 & 43.08 $\pm$ 0.09 & 3   \\
SDSS 0914$+$0507  & 0.301    &    44.84 $\pm$ 0.01           &    9.36 $\pm$ 0.10  &   2   &    42.85 $\pm$ 0.01 & 42.30 $\pm$ 0.08 & 5   \\
SDSS 0948$+$4335  & 0.226    &    44.76 $\pm$ 0.01           &    8.43 $\pm$ 0.03  &   0   &    42.44 $\pm$ 0.01 & $<$ 43.0         & 3   \\
SDSS 0954$+$0929  & 0.298    &    44.54 $\pm$ 0.01           &    8.56 $\pm$ 0.08  &   2   &    42.93 $\pm$ 0.01 & $<$ 41.60        & 5   \\  
       Ton 28   & 0.329      &    45.44 $\pm$ 0.01           &    8.03 $\pm$ 0.43  &   0   &    42.42 $\pm$ 0.02 & $<$ 42.34        & 2  \\
     4C 13.41   & 0.241      &    45.41 $\pm$ 0.01           &    9.54 $\pm$ 0.05  &   2   &    42.40 $\pm$ 0.02 & $<$ 41.41        & 2   \\ 
 HE 1029$-$1401   & 0.086    &    44.97 $\pm$ 0.05           &    8.70             &   0   &     \nodata         & 41.78 $\pm$ 0.12 & 4   \\
  PG 1049$-$005   & 0.359    &    45.59 $\pm$ 0.01           &    9.17 $\pm$ 0.03  &   0   &    43.47 $\pm$ 0.01 & $<$ 42.31        & 2   \\
     4C 61.20   & 0.421      &    45.26 $\pm$ 0.01           &    9.10 $\pm$ 0.09  &   1   &    43.34 $\pm$ 0.01 & $<$ 42.55        & 2   \\
     4C 10.30   & 0.422      &    44.78 $\pm$ 0.01           &    9.33 $\pm$ 0.05  &   2   &    42.75 $\pm$ 0.01 & $<$ 41.62        & 2   \\
     3C 249.1   & 0.313      &    45.42                      &    8.96             &   2   &    43.46            & 42.85 $\pm$ 0.02 & 1, 2   \\
SDSS 1131$+$2632  & 0.244    &    44.56 $\pm$ 0.02           &    9.08 $\pm$ 0.04  &   0   &    42.81 $\pm$ 0.01 & 42.11 $\pm$ 0.14 & 3   \\
SDSS 1139$+$4020  & 0.314    &    44.63 $\pm$ 0.01           &    8.21 $\pm$ 0.08  &   0   &    42.97 $\pm$ 0.01 & $<$ 41.61        & 5   \\ 
     4C 49.22   & 0.334      &    44.70 $\pm$ 0.01           &    8.50 $\pm$ 0.01  &   1   &    42.45 $\pm$ 0.01 & $<$ 41.81        & 2   \\
     GQ Comae   & 0.165      &    44.15 $\pm$ 0.01           &    8.36 $\pm$ 0.03  &   0   &    42.58 $\pm$ 0.01 & 40.19 $\pm$ 0.30 & 1, 2   \\
 PKS 1217$+$023   & 0.240    &    45.13 $\pm$ 0.01           &    8.87 $\pm$ 0.08  &   2   &    42.78 $\pm$ 0.01 & 41.82 $\pm$ 0.16 & 2, 3   \\
     4C 25.40   & 0.268      &    44.73 $\pm$ 0.01           &    8.86 $\pm$ 0.03  &   2   &    42.66 $\pm$ 0.01 & 42.61 $\pm$ 0.02 & 2   \\
       3C 273   & 0.158      &    46.04 $\pm$ 0.09           &    9.06             &   1   &    42.87 $\pm$ 0.10 & $<$ 41.73      & 1, 2, 3   \\
SDSS 1230$+$6621  & 0.184    &    44.53 $\pm$ 0.01           &    8.26 $\pm$ 0.03  &   0   &    42.66 $\pm$ 0.01 & 41.90 $\pm$ 0.15 & 3   \\ 
 HE 1228$+$0131   & 0.117    &    44.95 $\pm$ 0.09           &    8.29             &   0   &    42.23 $\pm$ 0.11 & $<$ 43.0         & 3   \\
SDSS 1230$+$1100  & 0.236    &    44.73 $\pm$ 0.01           &    8.60 $\pm$ 0.05  &   0   &    42.64 $\pm$ 0.01 & $<$ 43.0         & 3   \\
  PG 1307$+$085   & 0.154    &    44.80 $\pm$ 0.01           &    8.59 $\pm$ 0.05  &   0   &    42.69 $\pm$ 0.01 & $<$ 41.06        & 2   \\ 
  B2 1425$+$267   & 0.364    &    45.14 $\pm$ 0.02           &    9.66 $\pm$ 0.03  &   2   &    42.99 $\pm$ 0.01 & 42.37 $\pm$ 0.04 & 1, 2   \\
  PG 1427$+$480   & 0.221    &    44.62 $\pm$ 0.01           &    7.95 $\pm$ 0.03  &   0   &    42.52 $\pm$ 0.01 & $<$ 43.0         & 3   \\
SDSS 1444$+$0633  & 0.208    &    44.48 $\pm$ 0.01           &    8.33 $\pm$ 0.03  &   0   &    42.44 $\pm$ 0.01 & $<$ 43.0         & 3   \\
 HE 1453$-$0303   & 0.206    &    45.31 $\pm$ 0.12           &    8.59             &   0   &    42.54 $\pm$ 0.12 & $<$ 43.0         & 3  \\
SDSS 1507$+$4453  & 0.314    &    44.39 $\pm$ 0.01           &    7.34 $\pm$ 0.14  &   1   &    42.69 $\pm$ 0.01 & $<$ 41.66        & 5   \\      
SDSS 1507$+$1338  & 0.322    &    44.79 $\pm$ 0.01           &    9.67 $\pm$ 0.02  &   0   &    42.53 $\pm$ 0.01 & 41.96 $\pm$ 0.08 & 5   \\    
     4C 37.43   & 0.371      &    45.21 $\pm$ 0.01           &    9.37 $\pm$ 0.11  &   2   &    43.18 $\pm$ 0.01 & 43.02 $\pm$ 0.01 & 1,2    \\
  PG 1519$+$226   & 0.136    &    44.40 $\pm$ 0.01           &    7.90 $\pm$ 0.07  &   0   &    41.34 $\pm$ 0.09 & $<$ 43.0         & 1   \\
       OR 241   & 0.254      &    44.86 $\pm$ 0.01           &    8.44 $\pm$ 0.05  &   1   &    42.39 $\pm$ 0.01 & $<$ 41.85        & 2   \\
    3CR 323.1   & 0.265      &    45.20 $\pm$ 0.01           &    9.07 $\pm$ 0.03  &   2   &    42.99 $\pm$ 0.01 & 42.35 $\pm$ 0.13 & 1, 2, 3   \\   
     4C 11.50   & 0.436      &    44.99 $\pm$ 0.01           &    8.87 $\pm$ 0.08  &   2   &    43.23 $\pm$ 0.01 & 42.68 $\pm$ 0.03 & 2   \\
      Ton 256   & 0.131      &    44.51 $\pm$ 0.01           &    7.99 $\pm$ 0.02  &   1   &    43.00 $\pm$ 0.01 & 42.80 $\pm$ 0.12 & 1, 3   \\
SDSS 1655$+$2146  & 0.154    &    44.67 $\pm$ 0.01           &    8.28 $\pm$ 0.04  &   0   &    43.03 $\pm$ 0.01 & 43.03 $\pm$ 0.09 & 3   \\
  PG 1700$+$518   & 0.292    &    45.43 $\pm$ 0.09           &    8.98             &   0   &    42.63 $\pm$ 0.10 & 41.81 $\pm$ 0.18 & 2, 3   \\
      3CR 351   & 0.372      &    45.79 $\pm$ 0.01           &    9.81 $\pm$ 0.04  &   2   &    43.50 $\pm$ 0.01 & $<$ 42.37        & 2   \\
    II Zw 136   & 0.061      &    44.40 $\pm$ 0.10           &    8.27             &   0   &    41.95 $\pm$ 0.09 & 40.90 $\pm$ 0.32 & 1, 3   \\
 PKS 2135$-$147   & 0.200    &    44.97                      &    9.15             &   2   &    43.16            & 41.85 $\pm$ 0.04 & 2   \\
 HE 2152$-$0936   & 0.192    &    45.59 $\pm$ 0.10           &    8.76             &   0   &    42.17 $\pm$ 0.11 & $<$ 43.0         & 3   \\
 HE 2158$-$0107   & 0.213    &    44.89 $\pm$ 0.11           &    8.60             &   0   &    42.59 $\pm$ 0.09 & 42.32 $\pm$ 0.13 & 3   \\
 HE 2158$+$0115   & 0.160    &    44.21 $\pm$ 0.10           &    7.94             &   0   &    42.01 $\pm$ 0.09 & $<$ 43.0         & 3   \\
     4C 31.63   & 0.298      &    45.99                      &    8.54             &   1   &    42.92            & $<$ 42.08        & 1, 2   \\
  PG 2214$+$139   & 0.067    &    44.44 $\pm$ 0.11           &    8.64             &   0   &    41.81 $\pm$ 0.10 & $<$ 43.0         & 3   \\
  PG 2233$+$134   & 0.326    &    45.19 $\pm$ 0.01           &    8.57 $\pm$ 0.31  &   0   &    42.50 $\pm$ 0.07 & $<$ 40.58        & 2   \\
 PKS 2251$+$113   & 0.325    &    45.07                      &    9.15             &   2   &    42.98            & 42.51 $\pm$ 0.04 & 1, 2   \\
 HE 2307$-$0254   & 0.221    &    44.89 $\pm$ 0.10           &    8.39             &   0   &    42.00 $\pm$ 0.11 & $<$ 43.0         & 3  \\
     4C 09.72   & 0.433    &    45.17                        &    9.30             &   2   &    42.80            & $<$ 42.71        & 2   \\
 PKS 2349$-$014   & 0.174    &    44.77 $\pm$ 0.02           &    8.81 $\pm$ 0.03  &   2   &    42.45 $\pm$ 0.01 & 42.11 $\pm$ 0.11 & 3   \\
 HE 2353$-$0420   & 0.229    &    44.42 $\pm$ 0.10           &    8.12             &   0   &    42.58 $\pm$ 0.09 & 42.47 $\pm$ 0.12 & 3 
\enddata
\tablenotetext{a}{Radio flag. 0: RQQ, 1: FSRQ (flat-spectrum or core-dominant radio quasar), 2: SSRQ (steep-spectrum or lobe-dominant radio quasar).}
\tablerefs{(1) Boroson et al. 1985; (2) Stockton \& MacKenty 1987; (3) Husemann et al. 2008 (including the re-analyzed data given in Husemann et al. in prep.); 
  (4) Husemann et al. 2010; (5) this work.}
\end{deluxetable*}

We also collect the EELR measurements of 20 type-2 quasars (10 detections and 10 non-detections) presented by \citet{humphrey10} and \citet{villar-martin10,villar-martin11}.
Since they were selected from the SDSS type-2 quasars identified by \citet{zakamska03}, luminosity and EW of emission lines arising from the narrow-line region (NLR) 
as well as optical $ugriz$ magnitudes of the host galaxies are available in the archive.
We derive the rest-frame $g - i$ color and stellar mass of the host galaxies from the SDSS model magnitudes and spectroscopic redshifts, using a $k$-correction code 
developed by \citet{blanton07}.
The contributions of relatively strong [\ion{O}{2}] $\lambda$3727 and [\ion{O}{3}] $\lambda$4959, $\lambda$5007 lines from quasars are removed before the calculation 
by subtracting 
$\Delta m_X$ = $-2.5$ log (1 + $f_{X,Y}$ [EW$_Y$/$w_X$]) from $X$ ($u$, $g$, $r$, $i$, or $z$)-band magnitude if the line $Y$ is included in the band, 
where $f_{X,Y}$ is the filter transmission at the line $Y$ relative to the band average and $w_X$ is the band width.
We summarize the derived properties of the type-2 quasars in Table \ref{tab:gal_props}.
All but two sources, SDSS 1228$+$0050 and SDSS 1625$+$3106, are radio quiet \citep[SDSS 1625$+$3106 is in the transition range of radio-quiet and radio-loud objects;][]{villar-martin11}.
In addition, we use all the type-2 quasars in \citet{zakamska03} and $\sim$60,000 galaxies extracted from the SDSS galaxy sample at $z$ = 0.3 -- 0.6, the redshift range 
of the main sample (the type-2 quasars in Table \ref{tab:gal_props}), as control objects.
Their rest-frame color and stellar mass are calculated in the same way as for the objects in the main sample.

\begin{deluxetable*}{ccccccc}
\tablewidth{0pt}
\tablecaption{Nuclear and host galaxy properties of type-2 quasars with EELR measurements. \label{tab:gal_props}}
\tablehead{\colhead{} & \colhead{} & \colhead{log $L_{\rm N[O III]}$}  & \colhead{Rest-frame} & \colhead{Stellar mass} 
  & \colhead{} &\colhead{}\\
\colhead{Quasar} & \colhead{Redshift} & \colhead{(erg s$^{-1}$)\tablenotemark{a}}  & \colhead{$g - i$ (mag)\tablenotemark{b}} & \colhead{log $M_\star (M_\odot)$}
& \colhead{EELR\tablenotemark{c}} & \colhead{Ref.}}
\startdata
SDSS 0025$-$1040 & 0.303    & 42.32  &     0.75 &      11.0 &      $+1$  &  3\\
SDSS 0123$+$0044 & 0.399    & 42.72  &     1.14 &      10.9 &      $+1$  &  2\\
SDSS 0217$-$0013 & 0.344    & 42.34  &     0.93 &      11.1 &      $+1$  &  3\\
SDSS 0234$-$0745 & 0.310    & 42.36  &     0.91 &      10.5 &      $-1$  &  3\\
SDSS 0840$+$3838 & 0.313    & 42.21  &     1.13 &      11.4 &      $+1$  &  1\\
SDSS 0849$+$0150 & 0.376    & 41.65  &     1.01 &      10.9 &      $-1$  &  3\\
SDSS 0920$+$4531 & 0.402    & 42.63  &     0.91 &      11.1 &      $-1$  &  1\\
SDSS 0955$+$0346 & 0.421    & 42.19  &     1.24 &      11.2 &      $+1$  &  3\\
SDSS 1153$+$0326 & 0.575    & 43.20  &     0.62 &      11.2 &      $+1$  &  3\\
SDSS 1228$+$0050 & 0.575    & 42.87  &     0.85 &      11.0 &      $-1$  &  3\\
SDSS 1307$-$0214 & 0.425    & 42.51  &     1.09 &      11.4 &      $+1$  &  3\\
SDSS 1337$-$0128 & 0.329    & 42.31  &     0.90 &      11.3 &      $+1$  &  3\\
SDSS 1407$+$0217 & 0.309    & 42.49  &     0.86 &      10.5 &      $-1$  &  3\\
SDSS 1413$-$0142 & 0.380    & 42.84  &     1.18 &      11.0 &      $-1$  &  3\\
SDSS 1546$-$0005 & 0.383    & 41.77  &     1.18 &      11.0 &      $-1$  &  3\\
SDSS 1550$+$3950 & 0.347    & 43.05  &     0.60 &      11.3 &      $+1$  &  1\\
SDSS 1625$+$3106 & 0.379    & 42.60  &     1.35 &      11.1 &      $-1$  &  1\\
SDSS 1726$+$6021 & 0.333    & 42.16  &     0.96 &      10.5 &      $-1$  &  1\\
SDSS 1739$+$5442 & 0.384    & 42.01  &     1.24 &      11.1 &      $-1$  &  1\\
SDSS 2358$-$0009 & 0.402    & 42.91  &     0.88 &      11.1 &      $+1$  &  3
\enddata
\tablenotetext{a}{The measurement error is $<$0.01.}
\tablenotetext{b}{Typical error of a few tens of magnitude is expected, including the uncertainties in the photometry and $k$-correction.}
\tablenotetext{c}{EELR detection flag ($+1$: detected, $-1$: non-detected).}
\tablerefs{(1) Humphrey et al. 2010; (2) Villar-Mart{\'{\i}}n et al. 2010; (3) Villar-Mart{\'{\i}}n et al. 2011.}
\end{deluxetable*}

While the above samples of type-1 and type-2 objects are used below to probe the different aspects of quasars, we should keep in mind that
the two populations can be intrinsically different.
\citet{elitzur12} clarified this point by demonstrating that the two types of AGNs are drawn preferentially from the distribution of dust-torus
covering factors.
Since type-2 AGNs have dustier environments around them than type-1 counterparts, their host galaxies might tend to have higher star-formation
rates since (i) the presence of ample dust indicates active star formation producing them; and (ii) the negative feedback (suppression of star 
formation) by the AGN radiation might be weaker due to the dust obscuration.

\section{Emergence of EELR}

\subsection{Notes on New Sample}

\begin{figure}
\epsscale{1.0}
\plotone{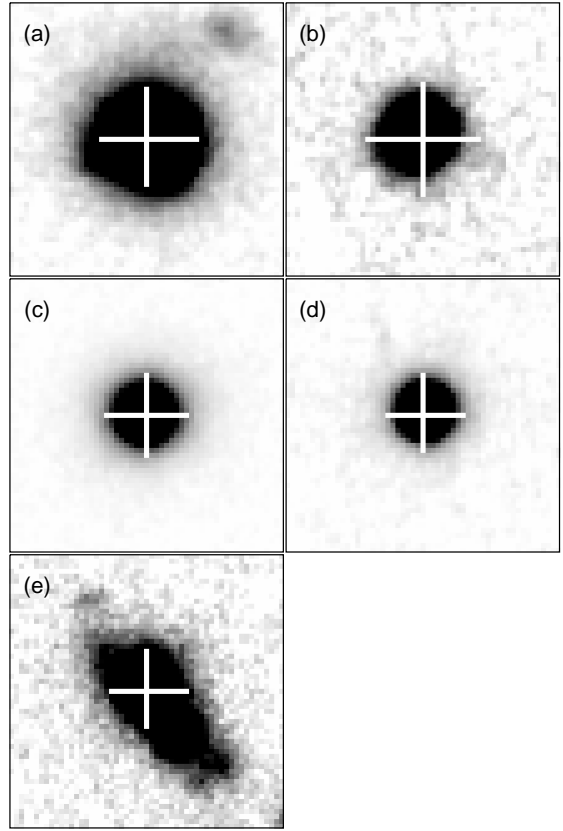}
\caption{Suprime-Cam [\ion{O}{3}] line images of (a) SDSS 0914$+$0507, (b) SDSS 0954$+$0929, (c) SDSS 1139$+$4020, (d) SDSS 1507$+$4453, and (e) SDSS 1507$+$1338.
  The grayscale represents linear flux levels between zero and the quasar peak values, while the white crosses indicate twice the FWHMs of the PSFs.
  The physical scale of the panels is 50 kpc on a side.
  North is up, east to the left.\label{lineimage}}
\end{figure}

The Suprime-Cam [\ion{O}{3}] line images of the newly observed quasars are presented in Figure \ref{lineimage}.
To aid visual inspection, the grayscale consistently represents linear flux levels between zero and quasar peak values while the white crosses indicate the PSF sizes 
(twice the full-widths at half-maximum [FWHMs]).
The most spectacular feature is found around SDSS 1507$+$1338.
The emission feature, extending across $\sim$50 kpc along the NE -- SW direction, is reminiscent of the ionization cones created by AGN radiation.
SDSS 0914$+$0507 has apparently extended emission too, with a knot at $\sim$35 kpc away from the center in the NW direction.
On the other hand, the radial profiles of SDSS 0954$+$0929, SDSS 1139$+$4020, and SDSS 1507$+$4453 are consistent with the PSFs and show no signs of EELR.
We measure the EELR luminosity in an annulus of inner radius 10 kpc and outer radius 30 kpc.
The wing components of the nuclear radiations are estimated from the PSFs and subtracted.
The incomplete transmission of the [\ion{O}{3}] lines through the NA$_{\rm 656}$ filter is also corrected using the filter transmission function.
We include the uncertainties in the continuum subtraction and in the correction for the wing contribution of the nuclear radiation, background sky noise, and
the possible flux variation assumed to be 10 \% due to the non-photometric sky condition, in the error budget of the derived luminosity.
The results are listed in the last column of Table \ref{tab:objprop}.
The above measurements support our visual inspection finding detectable EELR only in SDSS 0914$+$0507 and SDSS 1507$+$1338.
The nuclear [\ion{O}{3}] luminosity ($L_{\rm N[O\ III]}$) is also measured within a 3-arcsec aperture, and found to
agree with the SDSS spectroscopic measurements within $\Delta {\rm log} L_{\rm N[O\ III]} = \pm 0.1$.

\subsection{Link to AGN}

\begin{figure}
\epsscale{1.0}
\plotone{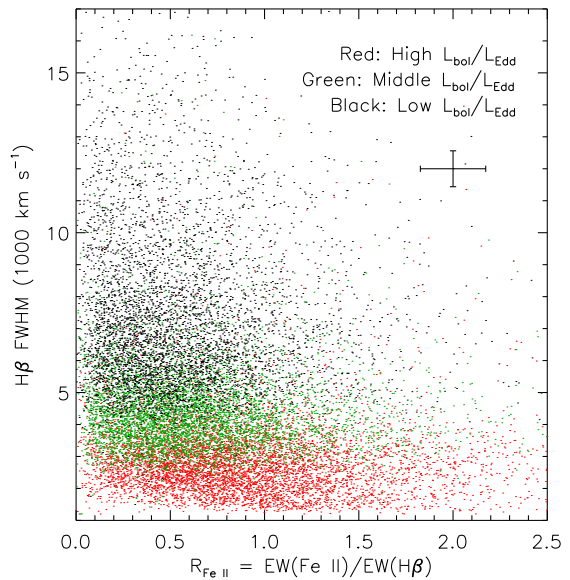}
\caption{H$\beta$ line width (FWHM) versus the relative \ion{Fe}{2} strength $R_{\rm Fe II}$ = EW({\ion{Fe}{2}})/EW(H$\beta$) of the SDSS quasars.
  The red, green, and black dots represent the objects with high ($-0.8 < $ log [$L_{\rm bol}$/$L_{\rm Edd}$]), middle 
  ($-1.2 < $ log [$L_{\rm bol}$/$L_{\rm Edd}$] $< -0.8$), and low (log [$L_{\rm bol}$/$L_{\rm Edd}$] $< -1.8$) values of Eddington ratio, respectively.
  The typical error of each data point is indicated at the top-right corner.
  \label{nuc_prop_p1}}
\end{figure}

As described in Section 1, the presence of EELR correlates with some AGN properties.
Quasars with larger [\ion{O}{3}] EW, broader and bumpier Balmer lines, fainter \ion{Fe}{2}, and SSRQ-like radio emission have 
higher possibility of accompanying EELR around them.
The correlation is fairly strong, as \citet{fu09} showed that if certain quasars are pre-selected based on their radio and nuclear [\ion{O}{3}] emissions,
$\sim$50 \% of them would turn out to have EELR as opposed to $\sim$20 \% with no a priori assumption.
Note that the above emissions arise from the different parts of AGN; broad Balmer lines and the \ion{Fe}{2} lines from parsec-scale BLR,
the [\ion{O}{3}] lines from kiloparsec-scale NLR, and radio and EELR emissions from more extended, galaxy-wide scales.
The correlation among them except for EELR has been well known since the seminal work by \citet{boroson92}.
With the principal component analysis technique, they showed that most of the variance in AGN spectral properties are contained in two sets of correlations.
A strong anti-correlation between the measures of [\ion{O}{3}] and \ion{Fe}{2} is the principal component 1 (PC 1), or eigenvector 1, with which Balmer 
line profile is also associated.
The general consensus is that PC 1 is driven mainly by Eddington ratio \citep[e.g.,][]{boroson02,sameshima11}.
In Figure \ref{nuc_prop_p1} we plot H$\beta$ line width versus the relative \ion{Fe}{2} strength $R_{\rm Fe II}$ $\equiv$ EW({\ion{Fe}{2}})/EW(H$\beta$)
of the SDSS quasars in \citet{shen11}.
The both measures constitute the four-dimensional eigenvector 1 (4DE1; other measures are the soft X-ray photon index and
\ion{C}{4} $\lambda$1549 line profile) space proposed by \citet{sulentic00}.
The anti-correlation between them is evident in Figure \ref{nuc_prop_p1}, which forms a sequence of Eddington ratio.
The data scatter reflects the orientation effects as well as the intrinsic variety of quasars, since one would observe larger line FWHM and $R_{\rm Fe II}$
\citep[\ion{Fe}{2} is emitted from the outermost part of BLR; e.g.,][]{matsuoka08}
as the viewing angle relative to the dust-torus plane decreases.
On the other hand, PC 2 links optical luminosity and \ion{He}{2} $\lambda$4686 line emission.
Its physical origin is thought to be mass-accretion rate to SMBHs because of its strong correlation with optical luminosity.
Therefore one would expect that the ratio of PC 1 and PC 2 is proportional to SMBH mass, which was actually proved by \citet{boroson02}.
Furthermore, \citet{laor00} found that the radio loudness is strongly related to SMBH mass as radio-loud/radio-quiet quasars are associated with most/least massive SMBHs.
The above arguments suggest that most of the AGN properties can be understood on the PC 1 - PC 2 plane.

\begin{figure*}
\epsscale{1.0}
\plotone{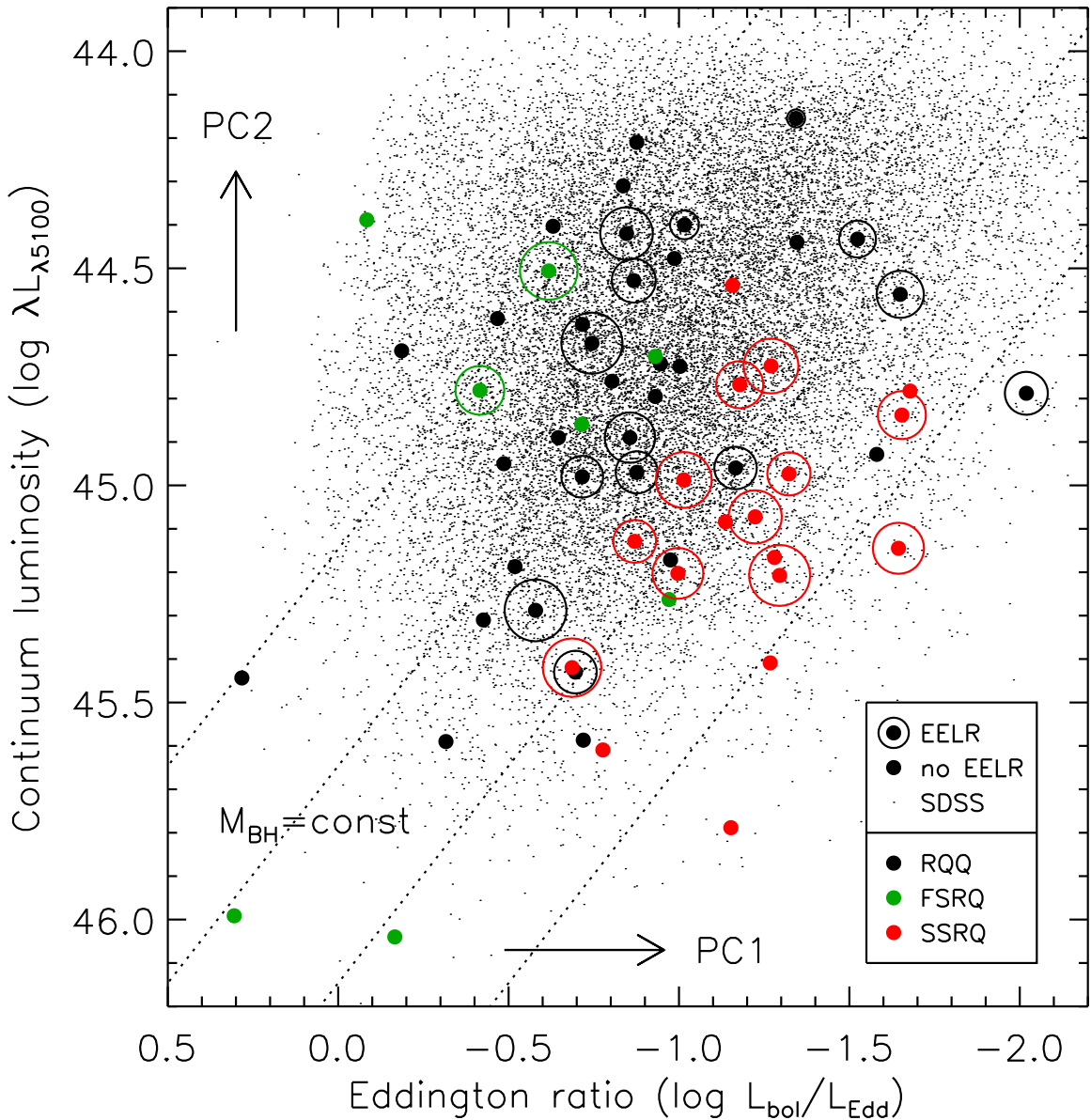}
\caption{Eddington ratio versus continuum luminosity of the compiled type-1 sample (filled circles) and the whole SDSS quasars (dots).
  RQQs, FSRQs, and SSRQs in the compiled sample are represented by black, green, and red colors, respectively.
  The quasars with detectable EELR are marked with the larger circles whose sizes are proportional to log $L_{\rm E[O III]}$.
  PC 1 and PC 2 increase along the horizontal and vertical axes on this plane, while SMBH mass increases from the top-left to the bottom-right corner.
  The lines of constant SMBH mass, log $M_{\rm BH}$ = 8.0, 8.5, 9.0, 9.5, are shown by the dotted lines.\label{nuc_prop_p4}}
\end{figure*}

\begin{figure}
\epsscale{1.1}
\plotone{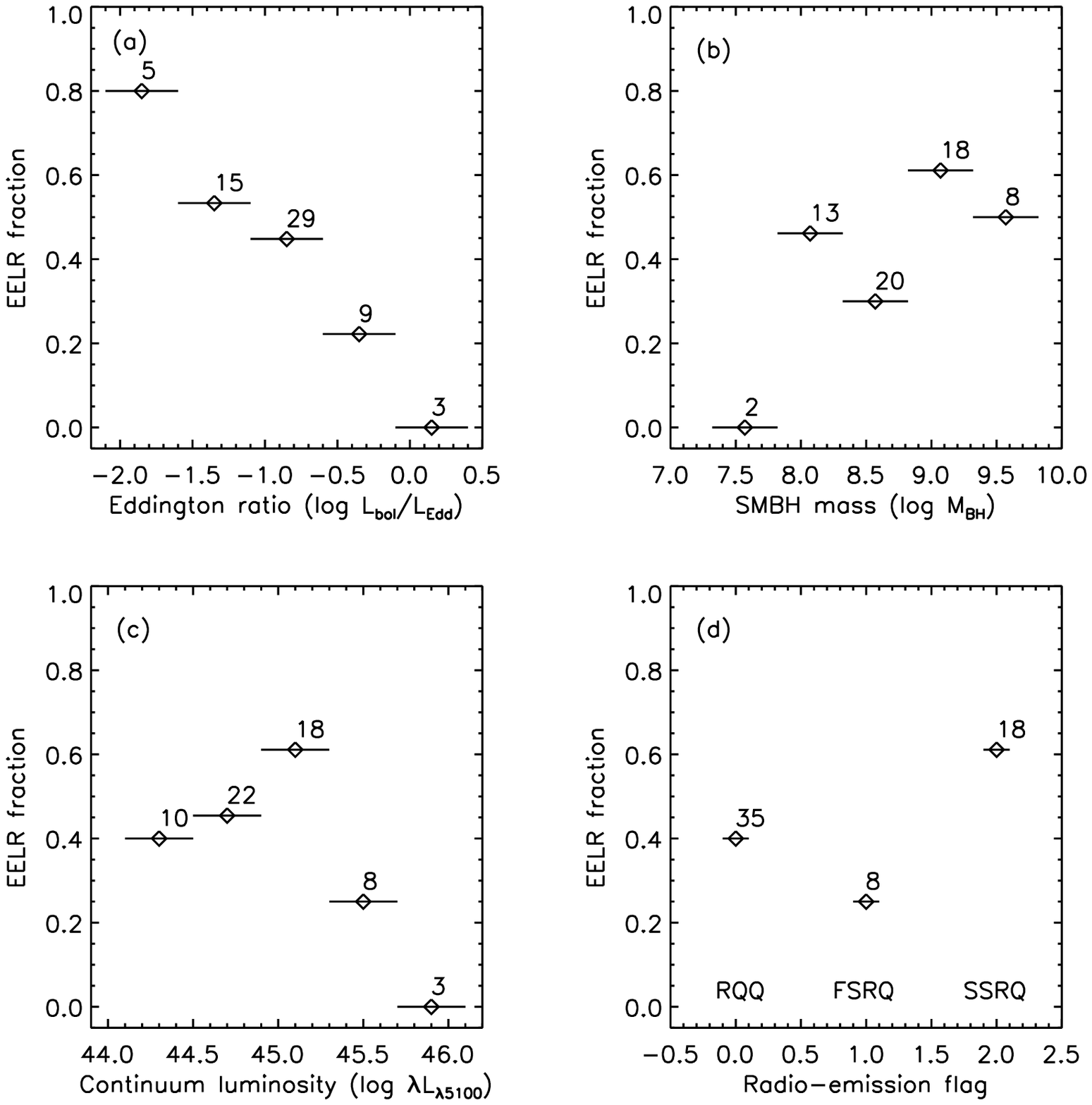}
\caption{Fraction of EELR detection as a function of Eddington ratio (panel a), SMBH mass (b), continuum luminosity (c), and radio-emission property (d).
  The fractions are calculated in the bins represented by the horizontal bars.
  The number of objects in each bin is also indicated.\label{nuc_prop_p9}}
\end{figure}

We plot Eddington ratio and continuum luminosity of the compiled type-1 sample, as well as the whole SDSS quasars in the \citet{shen11} catalog, 
in Figure \ref{nuc_prop_p4}.
First of all, we remark that the objects with EELR measurements cover the distribution of the whole SDSS quasars fairly uniformly on this plane.
It was not anticipated since the existing EELR observations are subject to strong selection biases individually.
We see the clear separation of radio-quiet and radio-loud sources as pointed out by \citet{laor00}, the latter being associated with most massive SMBHs.
In addition, SSRQs are observed to have higher values of Eddington ratio or SMBH mass than FSRQs.
We also find some correlations containing EELR.
They are shown more clearly in Figure \ref{nuc_prop_p9} where the fraction of EELR detection is plotted as a function of the nuclear properties.
The EELR fraction is strongly anti-correlated with Eddington ratio, and its weak correlation with SMBH mass may also be present.
The former relation is most likely the underlying basis of the correlation between EELR and PC 1.
While no clear trend is found with respect to radio emission, 
we do find that EELR luminosity is systematically higher in SSRQs 
than in the rest of the sample; the mean luminosity is $L_{\rm E[O III]}$ = $10^{42.4}$ and $10^{41.9}$ erg s$^{-1}$ for the SSRQs and RQQs, respectively.
The difference would be even larger when we consider the fact that the EELR luminosity of most SSRQs measured by \citet{stockton87} is underestimated since the emission
in a central 10-kpc aperture is excluded.
It explains why the EELR fraction around SSRQs is much higher than those around the other populations in the pioneering work of \citet{stockton87}.

\begin{figure*}
\epsscale{1.0}
\plotone{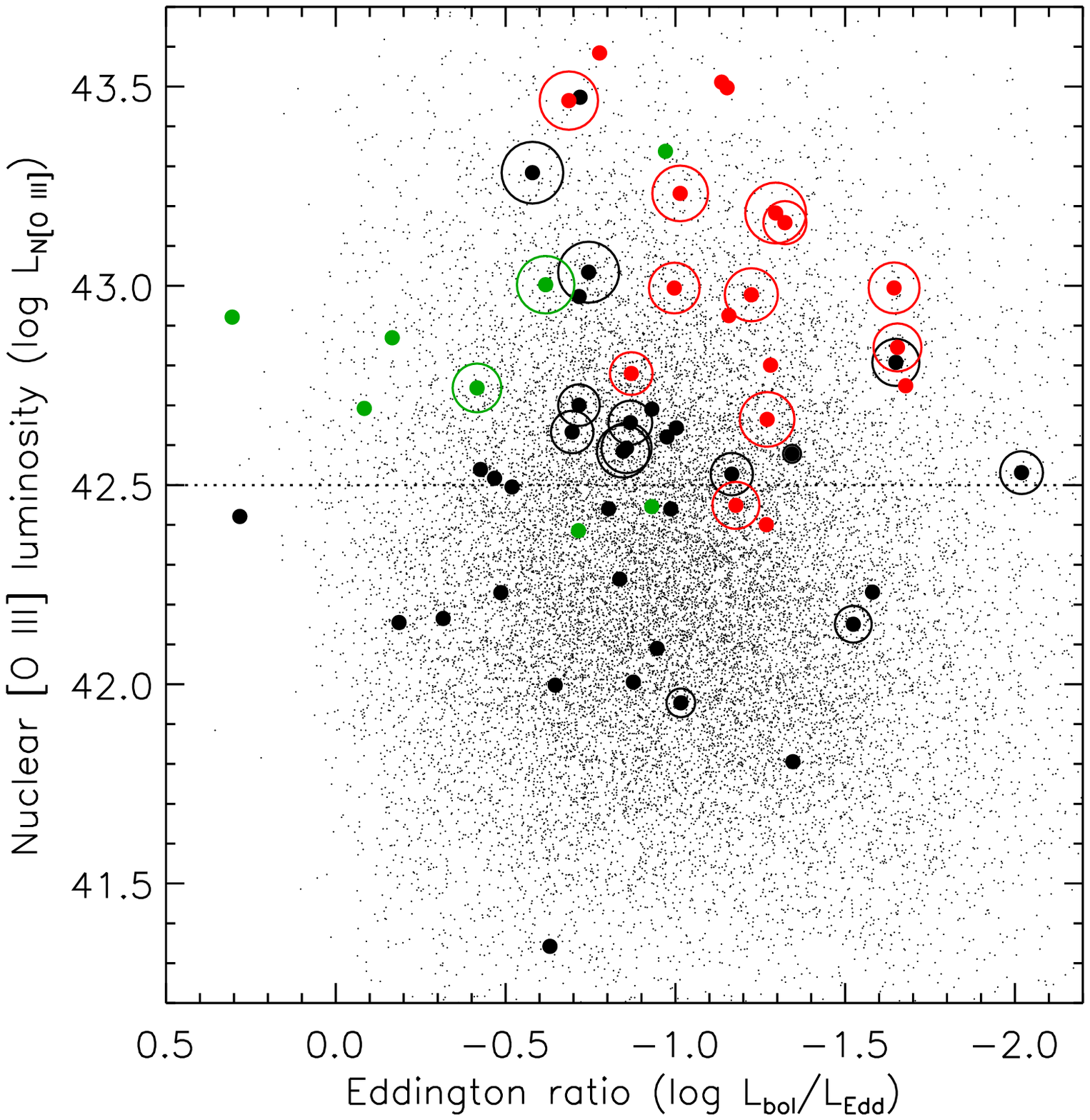}
\caption{Eddington ratio versus nuclear [\ion{O}{3}] line luminosity of the compiled type-1 sample and the whole SDSS quasars.
  The symbols are as in Figure \ref{nuc_prop_p4}.
  The dotted line represents $L_{\rm N[O III]}$ = $10^{42.5}$ erg s$^{-1}$.\label{nuc_prop_p6}}
\end{figure*}

One of the highest dependency of EELR is found on nuclear [\ion{O}{3}] luminosity, $L_{\rm N[O III]}$.
In Figure \ref{nuc_prop_p6} we show Eddington ratio versus $L_{\rm N[O III]}$ of the same samples.
It is clearly seen that the quasars with EELR measurements are biased to high $L_{\rm N[O III]}$ compared to the whole SDSS quasars.
Here we find that low-$L_{\rm N[O III]}$ quasars tend to lack EELR: only 3 out of 20 objects with $L_{\rm N[O III]} < 10^{42.5}$ erg s$^{-1}$ have detectable, faint EELR, 
while the EELR fraction is $>$ 50 \% at the higher $L_{\rm N[O III]}$.
This trend was reported previously for smaller samples \citep[e.g.,][]{stockton87}.
Figure \ref{nuc_prop_p6} also tells us that $L_{\rm N[O III]}$ does {\it not} strongly correlate with Eddington ratio, although it correlates with PC 1 \citep{boroson92}.
No clear correlation between the two measures indicates that PC 1 is not solely driven by Eddington ratio -- there should be another physical mechanism(s) that links nuclear 
[\ion{O}{3}] luminosity with other PC 1 constituents. 


What can we learn from these correlations?
While the clear anti-correlation between EELR and Eddington ratio is observed, it does not seem to reflect their direct relationship through the AGN radiation
since (i) the EELR fraction is highest for the AGNs with most {\it inefficient} gas accretion to SMBHs; 
(ii) nuclear [\ion{O}{3}] luminosity produced similarly by the AGN radiation shows no clear correlation with Eddington ratio;
(iii) a positive correlation between the EELR fraction and continuum luminosity is not observed.
Therefore it is more reasonable to consider the presence of an independent mechanism that leads to the high/low EELR fraction and low/high Eddington ratio at the same time.
The dependence of EELR on radio emission and (possibly) SMBH mass may also be related to this hidden mechanism.
We will come back to this issue further below.
The lack of simple relation between EELR and continuum luminosity (accretion rate) was also reported by \citet{fu09}, who suggested that the correlation between 
nuclear and extended [\ion{O}{3}] emissions, which is also evident in our larger sample, may be the sequence of the available gas amount rather than AGN emission 
characteristics.
The hypothesis that the gas availability is the primary determinant of EELR creation is supported by the probe of host galaxies in the following section.

\subsection{Link to Host Galaxies}

\begin{figure}
\epsscale{1.2}
\plotone{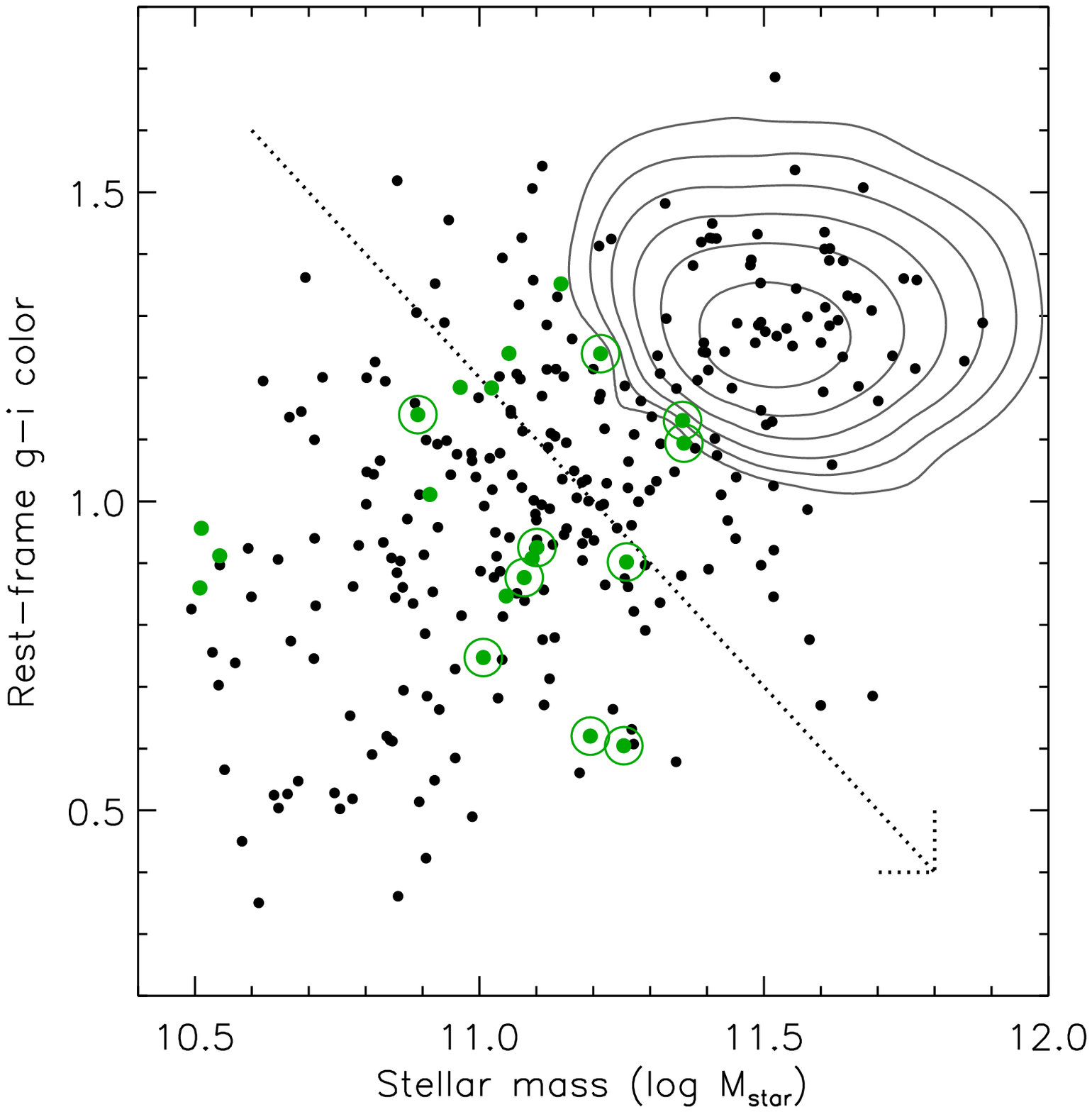}
\caption{Stellar mass versus rest-frame $g - i$ color of the galaxies hosting type-2 quasars with EELR measurements (green circles).
  The objects with detectable EELR are marked with the larger circles.
  The type-2 quasars of \citet{zakamska03} at $z$ = 0.3 -- 0.6 are represented by the black circles, while
  the distribution of $\sim$60,000 SDSS galaxies in the same redshift range is shown by the contours (the contour levels are logarithmic).
  The arrow represents the 'massive-blue'-ness indicator $\mathcal{I}_{\rm mb}$ used in Figure \ref{gal_prop_p2}.
\label{gal_prop_p1}}
\end{figure}

\begin{figure}
\epsscale{1.0}
\plotone{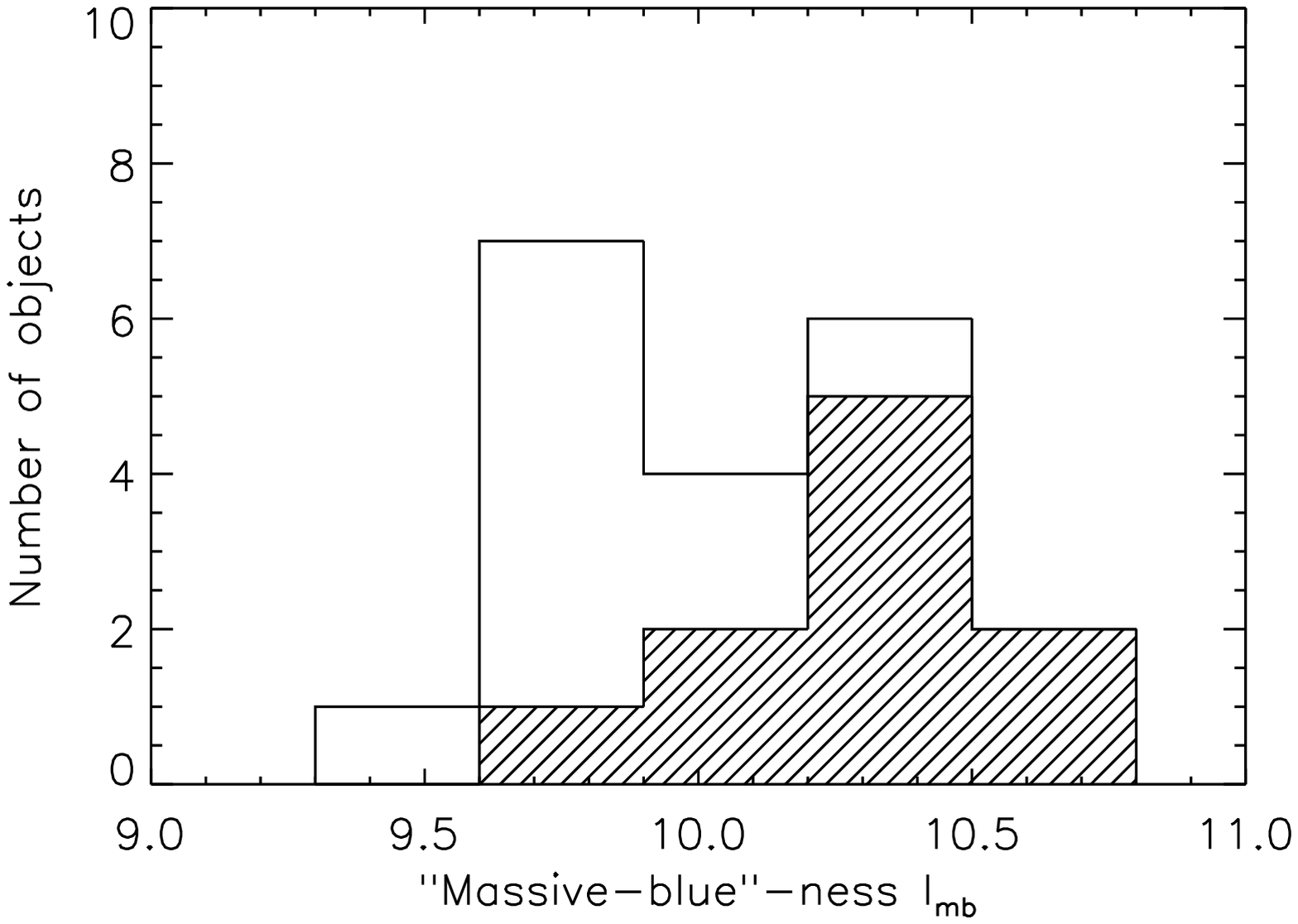}
\caption{Number distribution of the type-2 quasars with EELR measurements as a function of the 'massive-blue'-ness, $\mathcal{I}_{\rm mb}$ = log M$_{\star}$ $-$ ($g - i$).
  The hatched area is for the EELR quasars.\label{gal_prop_p2}}
\end{figure}

If EELR represents a certain stage of galaxy evolution, as the AGN feedback scenario implies, their presence may be related to stellar-population properties of host galaxies.
Here an investigation of such a relation is presented for the first time, thanks to the recent measurements of EELR around SDSS type-2 quasars at $z$ = 0.3 -- 0.6
\citep{humphrey10,villar-martin10,villar-martin11}.
In Figure \ref{gal_prop_p1}, we plot stellar mass versus rest-frame $g - i$ color of the type-2 quasars as well as the control samples
defined in Section \ref{sec:datacomp}.
Most SDSS galaxies at these relatively high redshifts are luminous red galaxies (LRGs), forming a clump at the massive red end on this color - stellar mass diagram
(we could see the red sequence rather than a clump if the SDSS observations were deep enough to detect fainter galaxies).
The type-2 quasars are bluer than the LRGs \citep[see also,][]{zakamska03} but are redder than star-forming galaxies which form the "blue cloud" around
$g - i \sim 0.6$ in the local universe \citep[e.g.,][]{gavazzi10}.
It is consistent with the result of \citet{schawinski10} who found that local AGNs reside in the "green valley" between the blue cloud and the red sequence
on the color - stellar mass diagram.
Interestingly, we find that the EELR quasars are associated with bluer and more massive galaxies compared to the non-EELR counterparts.
In order to show this trend more clearly, we define a 'massive-blue'-ness indicator, $\mathcal{I}_{\rm mb}$ $\equiv$ log M$_{\star}$ $-$ ($g - i$), which increases diagonally
toward the bottom-right corner of Figure \ref{gal_prop_p1} as shown by the arrow.
The number distribution of the type-2 quasars with EELR measurements as a function of $\mathcal{I}_{\rm mb}$ is plotted in Figure \ref{gal_prop_p2}.
It shows a clear, monotonic increase in the fraction of EELR quasars as $\mathcal{I}_{\rm mb}$ increases: only 1/8 of the lowest-$\mathcal{I}_{\rm mb}$ quasars has EELR,
while the fraction rises to 7/8 in the highest-$\mathcal{I}_{\rm mb}$ counterparts.

What is the origin of this dichotomy?
Blue colors point to current or recent star-formation activity, which in turn indicates the presence of a certain amount of gas in the galaxies.
Based on the velocity structure and mass of EELR gas around mostly SSRQs, as well as the correlation between EELR and radio emission, \citet{fu09} suggested 
that EELR is formed in galaxy minor merger which is accompanied by production of a radio jet associated to AGN.
A similar conclusion was reached by \citet{husemann10} for a RQQ.
\citet{villar-martin11} found that the majority of the type-2, radio-quiet EELR quasars show the signs of galaxy merger or interaction
while none of the non-EELR counterparts show such signs.
These observations suggest that the gas may be brought in by the merging companions.
The merger/interaction possibly activates the observed star formation as well as AGN that ionizes the gas and creates EELR, since
the interstellar gas of merging galaxies can be effectively brought into the nuclear regions with rapid loss of the angular momentum \citep[e.g.,][]{barnes96}.
The fact that EELR is preferentially found in gas-rich, massive blue galaxies is consistent with both the above scenario and the previous arguments
that the presence of EELR is regulated by the amount of available gas.
Furthermore, massive galaxies are known to be the dominant hosts of radio-loud AGNs \citep[e.g.,][]{floyd04,best05}.
It naturally explains the EELR - radio emission connection described above, and can also provide the "hidden" mechanism behind the anti-correlation between EELR and
Eddington ratio since the majority of radio-loud AGNs are powered by inefficiently-accreting, most massive SMBHs (see Figure \ref{nuc_prop_p4}).


The recent kinematic observations have revealed the presence of galaxy-wide, disturbed gas in quasar host galaxies \citep[e.g.,][]{fu09,greene11}.
While only a small fraction of the gas is generally found to have high enough velocity to escape the galaxies, \citet{greene11} showed that
these estimates are in fact lower limits and may be consistent with expectations of recent AGN feedback models.
If the feedback scenario is correct, the gas responsible for EELR and star formation (hence blue colors of galaxies) would soon be
swept away by the AGN-driven, galactic winds. 
In this regard, it may be noteworthy that none of the most luminous (log $\lambda L_{\lambda5100}$ $\ge$ 45.5) type-1 quasars in our sample has EELR 
as shown in Figure \ref{nuc_prop_p4}.
It is tempting to raise the possibility of the gas stripping, but we have to be aware that (i) the relatively-high Eddington ratio for these objects may be
responsible for the deficit of EELR; (ii) the bright central radiation can prevent us from detecting EELR in the most luminous sources.
More data is needed for further inspection on this issue.
While the process discussed above corresponds to the so-called "quasar-mode" feedback, it is interesting to point out that a clear interaction between radio 
jets and EELR clouds with high velocity dispersion is observed in SSRQs \citep{fu09}, which invokes the another, "radio mode" of the feedback.

In the nearby universe, \citet{keel11} investigated the properties of Seyfert galaxies with EELR, dominated by type-2 objects.
While their morphology supports the merger origin of EELR, they have relatively quiescent kinematics without a sign of
high-velocity gas leaving the galaxies seen in higher-luminosity AGNs. 
It may simply indicate that creation of the powerful galactic winds requires quasar-class energy input into the interstellar medium,
hence the relation of EELR to AGNs may be different at high and low luminosity.

\section{Summary and Conclusions}

A comprehensive analysis of EELR phenomenon around quasars is presented in this paper.
We compile the past EELR measurements for type-1 \citep{boroson85,stockton87,husemann08} and type-2 \citep{humphrey10,villar-martin10,villar-martin11} quasars,
and combine them with the new observation of five quasars at $z \sim 0.3$.
The new observation was carried out using the Subaru/Suprime-Cam with the narrow-band NA$_{656}$ filter (corresponding to the redshifted [\ion{O}{3}] $\lambda$5007),
and led to the discovery of EELR around two sources, SDSS 091401.75$+$050750.6 and SDSS 150752.66$+$133844.5.
The observation effectively complements the existing measurements on the AGN principal-component (PC) plane.
The properties of associated AGN (SMBH mass, continuum and line luminosity, and radio characteristics) and host galaxies (stellar mass and rest-frame color) 
are collected and calculated from the SDSS and other observations.

We find that EELR anti-correlates with Eddington ratio, which is most likely the underlying basis of the known correlations between EELR and the PC 1 (eigenvector 1) constituents.
EELR and nuclear [\ion{O}{3}] emissions are also highly correlated.
We argue that the primary determinant of EELR creation is the gas availability, rather than AGN emission characteristics.
The recent observations \citep{fu09, husemann10, villar-martin11} revealed that a significant fraction of EELR galaxies show the signs of recent minor merger,
which may have brought in the gas and activated star formation as well as AGN that creates EELR.
In line with the above arguments, we find that EELR is preferentially associated with gas-rich, massive blue galaxies.
It may explain the whole set of observed correlations containing EELR, since massive galaxies are the dominant hosts of radio-loud AGNs characterized with low
Eddington ratio and high SMBH mass.

\begin{figure*}
\epsscale{1.0}
\plotone{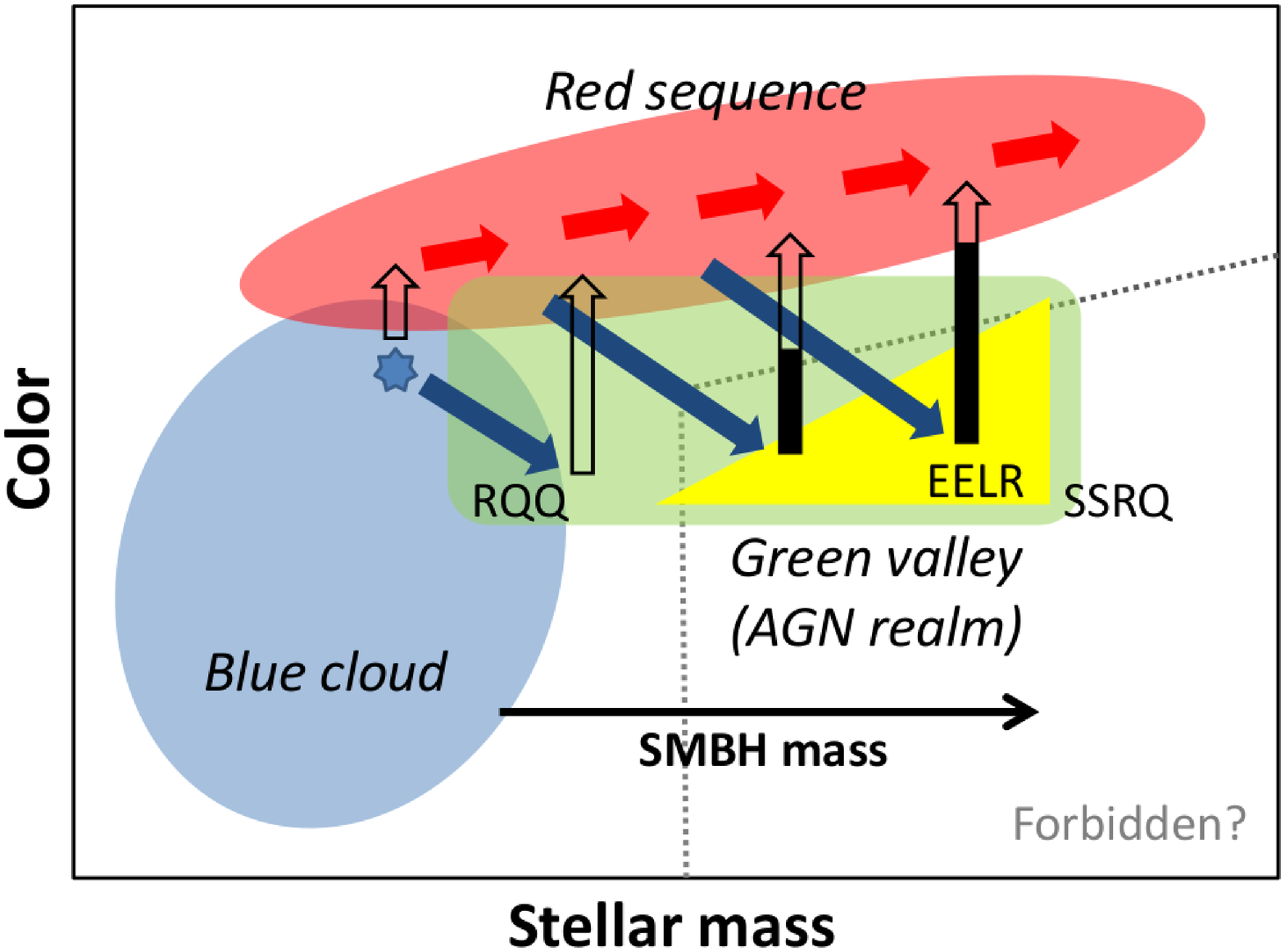}
\caption{Schematic view of galaxy color - stellar mass diagram, or equivalently CMD.
  The blue, red, and green regions represent the blue cloud of star-forming galaxies, the red sequence of quiescent galaxies, and the green valley occupied by
  AGN host galaxies, respectively.
  The implied location of EELR galaxies is also shown by the yellow triangle.
  The bulge mass - SMBH mass correlation tells us that SMBH mass increases toward right, hence the habitats of RQQs and SSRQs are separated in the green
  valley as indicated.
  The lower-right corner of this diagram enclosed by the dotted lines seems to be the "forbidden" area for observed galaxies.
  The seven-rayed star marks the starting point of the galaxy evolution discussed in the text.
  The blue and red arrows represent wet and dry merger, respectively, while the open arrows represent the stable color change due to star-formation fading.
  The filled parts of the open arrows show the contribution of the violent star-formation quenching by the AGN feedback.
  \label{picture}}
\end{figure*}

The hierarchical structure formation models based on the $\Lambda$CDM theory predict that galaxies are assembled through a sequence of mergers of smaller building blocks.
In Figure \ref{picture} we show a schematic view of the color - stellar mass diagram, or equivalently the color - magnitude diagram (CMD), of observed galaxies.
Once a galaxy is grown up to the massive tip of the blue cloud (represented by the seven-rayed star in the figure), it has two ways to evolve further.
If no merger happens subsequently for sufficient time, the galaxy would migrate into the low-mass end of the red sequence as the star formation fades out.
It can then move up the red sequence toward the massive end through dry merger with gas-poor galaxies \citep[see also,][]{faber07}.
On the other hand, wet merger of the galaxy with gas-rich galaxies at any stage of the above evolution could push the galaxy into the green valley 
where AGN fraction is the highest.
If the galaxy (and hence the central SMBH, assuming the positive bulge mass - SMBH mass correlation) is massive enough and the merging events involve ample supply of gas 
for triggering vigorous star formation, the galaxy would be driven to the massive blue realm at the extreme corner of the valley.
This event would also activate AGN, which ionizes the gas and creates luminous EELR.
The galaxy-wide energy injection by the AGN radiation might then blow away the gas and quench the star formation, pushing the galaxy back to the red sequence.
While this last argument is less certain due to the lack of evidence about the energy injection process, 
recent observations have started to reveal the presence of AGN-driven, galaxy-scale outflows which may be quenching the star formation there.
\citep[e.g.,][]{greene11,rupke11,cano-diaz11}.

The above AGN feedback scenario gives a solution to the major problems of the classical hierarchical galaxy formation models, i.e., the overproduction of 
massive galaxies and the inverted 
color - magnitude - morphology relation (see Section 1) by suppressing star formation in massive galaxies.
However, many details of the process are still unknown and to be debated.
What is observationally clear is that EELR is found at the exact region on the CMD where the AGN feedback is expected to work -- the massive blue corner 
that seems to be the forbidden area for observed galaxies.
It provides a piece of evidence for the presence of the AGN feedback, which may be playing a leading role in the co-evolution of galaxies and central SMBHs.

\acknowledgments

The author is grateful to B. Husemann for kindly providing the unpublished data on their re-analysis of the EELR measurements.
The anonymous referee has provided many suggestive comments to improve the manuscript.
A great assistance was provided for the Subaru/Suprime-cam observation by S. Masaki, R. Asano, and the staff at the Hawaii observatory, NAOJ.
This work was supported by Grant-in-Aid for Young Scientists (22684005) and the Global COE Program of Nagoya University 
"Quest for Fundamental Principles in the Universe" from JSPS and MEXT of Japan.

\end{document}